\documentclass[epj]{svjour}
\usepackage{graphics}
\begin{document}
\title{{\boldmath $\gamma$}-RADIATION OF EXCITED NUCLEAR DISCRETE 
LEVELS  IN PERIPHERAL
HEAVY ION COLLISIONS}
\author{V.L.Korotkikh \and K.A.Chikin
}                     % Do not remove
\institute{Scobeltsyn Institute of Nuclear Physics, Moscow State University, 119899 Moscow, Russia. E-mail: vlk\symbol{64}lav01.sinp.msu.ru}
\date{Received: 14 December 2001}
\abstract{
    A new process of a nuclear excitation to  discrete states in peripheral heavy ion collisions is studied. High energy photons are emitted by the exited nuclei with energies up to a few tens GeV at  angles of a few hundred microradians with respect to the beam direction.  We show that a  two stage process where an electron - positron pair is produced by virtual photons emitted by  nuclei and then the electron or positron excites the nucleus, has  large cross section. It is equal to about 5 b for CaCa collisions. On one hand it produces a significant $\gamma$-rays background  in the nuclear fragmentation region but on the other hand it could be used to  monitoring the nuclear beam intensity at LHC. These secondary nuclear photons could be a good signal for triggering  peripheral nuclear  collisions.
 % end of PACS codes
} %end of abstract
\maketitle
\authorrunning {V.L.Korotkikh \and K.A.Chikin}
\titlerunning {{\boldmath $\gamma$}-RADIATION OF EXCITED NUCLEAR DISCRETE 
LEVELS  IN PERIPHERAL
HEAVY ION COLLISIONS}

\section{Introduction}
\label{intro}
Two photon physics calls for attention. It includes the effects of
coherent photon interactions for nuclear peripheral collisions. Such processes takes place at large distances between nuclei,  $b > 2 R_A$, where
$b$ is the impact parameter. The interactions are therefore due to  long range 
electromagnetic processes. The cross section is proportional  to $(\alpha Z)^4$ for $Z_1 = Z_2 = Z$, where $Z$ is a charge of each ion 
$A_Z$. The ions are on essentially undisturbed and fly in the
beam direction.

There are many theoretical studies of two virtual photon
fusion to a particle of mass $M$ in  peripheral $AA$ collisions:
\begin{equation}
\label{1}
\gamma^* + \gamma^* \to M.
\end{equation} 
Photons fusion can produce particles from $\mu^\pm$, $\tau^\pm$ leptons
to Higgs bosons~\cite{1,2,3,4,5}.  Research in this field can yield  fundamental results. A full list of references can be found in the recent reviews~\cite{1,2,3}.

We considered three competitive processes in our previous work~\cite{5} in which $\pi^0$-mesons are produced in peripheral $AA$ collisions. Virtual photon fusion  
\begin{equation}
\label{1a}
\gamma^* + \gamma^* \to \pi^0
\end{equation}
can give  $\pi^0$-mesons  in the central rapidity region and at $p_T$ less than  $ 75$ MeV. The photo-disintegration of nucleus 
\begin{equation}
\label{1b}
\gamma^* + A_Z \to \pi^0 + X
\end{equation}
also produces  $\pi^{0}$-mesons with large amount of nuclear fragments in
the fragmentation region~\cite{6,7}. Hadronic
interactions in the tails of the nuclear density at $b > 2R_A$, 
\begin{equation}
\label{1c}
A_Z+A_Z \to \pi^0 + X
\end{equation}
produces $\pi^0$s in the whole rapidity region together with the
nuclear fragments and other particles in the fragmentation region. 
We demonstrated~\cite{5}
that to distinguish the fusion process (\ref{1a}) from processes
(\ref{1b}) and (\ref{1c}) it is necessary to choose a  strong $p_T$ cut, $p_T < 75$ MeV,
for all produced particles. Additionally it is necessary to have an efficient
trigger to distinguish photon-photon events from hadronic ones. G.Baur et
al.~\cite{3} suggested  measuring  intact nuclei after the interaction. Evidently it
is impossible for CMS. The nuclei at such low $p_T$ will emit
in very small solid angle, $\theta_A \leq 1$ $\mu$rad, at LHC energies and fly into the beam pipe.  We are interested in the fusion process (\ref{1a}) where both
nuclei preserve their $A$ and $Z$. It is the necessary to remove the  contributions in Eqs. (\ref{1b}) and (\ref{1c}) by the help of the veto-detectors on
nuclear fragments. 

The separation of peripheral and central $AA$ collisions is a hard experimental problem.
There are two suggestions to select peripheral events: the correlation between  $b$
 and total transverse energy  $E_T$~\cite{24Emel} or the correlation of $b$ and particle multiplicity~\cite{25Nyst}. But there are no ``event by event'' criteria for peripheral collision selection. 

Two photon fusion (\ref{1}) is accompanied by other electromagnetic processes which have to be considered if we want to obtain a clear signature of the process (\ref{1}).

The pure electromagnetic processes of the electron-positron pair
production~\cite{8,9,10,11} and bremsstrahlung photons~\cite{12,13,14,15} are
also
studied. These processes are considered as a possible background contribution.
The cross section of $e^+ e^-$ pair production is huge, 220
Kb for PbPb and 1.5 Kb for CaCa collisions at LHC energies. 
We will use the fact of the huge $e^+ e^-$ cross section below.

The physical program of LHC heavy ions includes the nucleus Pb, Nb, Ar and O.
Our calculations for CaCa are close to ArAr collisions. We choose Ca
because this nucleus is studied very well, the electromagnetic form factors
are well-known from experiments. We have no free parameters in our calculations.

The direct
bremsstrahlung from the heavy ion 
\begin{equation}
\label{2}
A_Z + A_Z \to A_Z + A_Z + \gamma_{BS}' 
\end{equation}
is small~\cite{12,13,14} with a cross section  proportional to  $Z^6
\alpha^3/M_A^2$. Ref.~\cite{15} suggested measuring the bremsstrahlung
photons coming from  electrons and positrons produced in
peripheral ion collisions, 
\begin{equation}
\label{3}
A_Z+A_Z \to A_Z+A_Z+e^+e^- + \gamma_{BS}'. 
\end{equation} 
The cross section is proportional to $Z^4 \alpha^5/m_e^2$,  larger than the cross section of process (\ref{2}). We used the
result of Ref.~\cite{15} for PbPb  and found about 10 b for
CaCa at LHC energies.  The energies of bremsstrahlung photons are small,
$E_{\gamma'_{BS}} \leq 3$ MeV, and the angles are near $\theta_{\gamma'_{BS}} =
1^\circ$. Such low energy photons are impossible to measure. So they are not a good 
 signature of peripheral $AA$ collisions.

In the present work we study a new process in which a nucleus is excited 
to  discrete levels by an 
electron
(positron) produced in electromagnetic interaction between nuclei
\begin{equation}
\label{4}
A_Z + A_Z \to A_Z^* + A_Z + e^+ e^- .
\end{equation}

The excited nucleus $A_Z^*$ radiates the secondary \\
photon $\gamma'$
\begin{equation}
\label{5}
A_Z^* \to A_Z + \gamma'. 
\end{equation}
As a rule, the  nuclear excitation energy is  a few MeV~\cite{16}. The
energies $E_{\gamma'}^0$ of photon in the nucleus rest system are in MeV scale.
They will be equal to about 10 GeV or more in the laboratory system  at LHC energies. 
The polar angle $\theta_{\gamma'}$ of the secondary photon is about a
few hundred $\mu$rad. It allows for a favourable conditions of the photon
measurement and using them as a  trigger for peripheral collisions.  
The process (\ref{5}) was suggested as a possible explanation of
the high energy ($E_\gamma \geq 10^{12}$ eV) cosmic photon
spectrum~\cite{Balashov}. 

There are other processes which could produce  secondary photons.
Mesons are produced by fusion process (\ref{1}) and decay to photons 
such as $\pi^0 \to 2\gamma$.
Neutral pions can also be produced  by  double Pomeron($IP$)-exchange~\cite{17} in $AA$ collisions,
\begin{equation}
\label{5a}
 IP +  IP \to \pi^0 + X.
\end{equation}
The calculations of the PHOJET event generator ~\cite{18} for
Pomeron exchange give an average transverse momentum of about 450 MeV for pions.
The calculations in Ref.~\cite{Roldao} show that cross section of the double Pomeron($IP$)-exchange is smaller than fusion process (\ref{1}) for heavy ions and is comparable for medium
nuclei.    

There are other competitive processes which  also produced secondary
photons. Such as nuclear excitation, 
\begin{equation}
\label{6}
A_Z + A_Z \to A_Z^* + A_Z,~~~~~A_Z^* \to \gamma' + A_Z, 
\end{equation}
in  peripheral collisions at $b > 2R$. We will show that the
cross section is smaller than process (\ref{4}). It is also possible 
to excite one nucleus by a virtual photon from other nucleus, 
\begin{equation}
\label{7}
\gamma^* + A_Z \to A_Z^*,~~~~~~A_Z^* \to \gamma' + A_Z. 
\end{equation}
The cross section of this process is the same or smaller
than (\ref{6}) for intermediate mass ions 
with $Z_A \approx  20$.  

Other processes which are also a  source of secondary photons are the
excitation of nuclei to the continuum  with  cascade transitions to
lower nuclear states that emit photons. Such processes have a
large probability to decay, $A^* \to (A-4) +~^4{\mbox He}$, $A^* \to
(A-1) + {\mbox N}$, if the excitation energy is higher than particle decay
threshold~\cite{16}. It might be possible to remove  these decays
by vetoing  charged particles. We also
need  to exclude the secondary photon contribution from nuclear fragments in
 photo-disintegration reactions 
\begin{equation}
\label{8}
\gamma^* + A_Z \to A_1^* + A_2^* + \ldots,~~~~~A_k^* \to \gamma + A_k.
\end{equation}
Our estimates, based on Ref.~\cite{6} using RELDIS, show that the nuclear fragments scatter in small angles $\theta \leq 45$ $\mu$rad. These fragments do not hit the
central detectors but can be removed either by the quadruple lens 
or by the Roman pots of TOTEM~\cite{19}. 

The TOTEM detector will measure  protons   at small $t$, $0.02 < |t| < 0.7$ (GeV/c)$^2$,
corresponding to small polar angles, $20 < \theta < 120$ $\mu$rad,
with  respect to the
beam direction. 
We will  shown that cross section of (\ref{4}) is large, $\sigma
\approx 5$ barn, so that secondary photons  hit the Roman pots with high rates and can be a significant background 
resulting in degradation of the detector.

There are two ways to use the secondary photons produced by 
 process ~(\ref{4}). 
One  possibility is
to use them for monitoring the nuclear beam intensity . At RHIC
a neutron zero degree calorimeter (ZDC)~\cite{20} is used.
The beam crossing angle at RHIC is 14 mrad and the ZDC
measure neutrons in the region $\Delta \theta = 6$ mrad. The LHC beam crossing
angle  is very small,  300 $\mu$rad. As yet, a
nuclear beam intensity monitor has not developed. The
possibility of using (\ref{4}) to monitor of the nuclear beam at LHC is
important.  

The second application concerns  triggering 
peripheral nucleus-nucleus collisions in processes such as 
\begin{equation}
\label{9}
A_Z A_Z \to A_Z^* A_Z + e^+e^- + M, ~~~~~A_Z^* \to \gamma' + A_Z;
\end{equation}
\begin{equation}
\label{10}
A_Z A_Z \to A_Z^* A_Z + M, ~~~~~~~~~~~~~~~A_Z^* \to \gamma' + A_Z, 
\end{equation}
where $M$ is the produced particle system. The total trigger requirements
include the following features: a signal in the central rapidity region from
$\gamma^* \gamma^* \to M$ events, the absence of charge particles
in the nuclear fragmentation region and a photon signal in the Roman
pots. We do not calculate these processes here but
study the angular and energy distributions of secondary
photons in decay (\ref{5}) for  process (\ref{4}). Note that these distributions should be similar to those of processes (\ref{9}) and (\ref{10}).

\section{Formalism} 
\subsection{Secondary nuclear $\gamma$-radiation} 

We study photon radiation of the relativistic nucleus   from discrete level 
\begin{equation}
\label{11}
A^*(J^P,E_\gamma^0) \to A + \gamma{'}.
\end{equation}
The secondary photon  $\gamma{'}$ flies off at angle $\theta_\gamma{'}$ and
energy $E_\gamma^0$ in the nucleus rest (NR) system. 
If a nucleus has Lorenz factor $\gamma_A$ then the secondary photon energy is 
\begin{equation}
\label{12}
E_\gamma = \gamma_A E_\gamma^0 (1+\cos \theta_\gamma{'})
\end{equation}
in the laboratory system (LS) so that ~$0 \leq E_\gamma \leq 2 \gamma_A
E_\gamma ^0$ and
\begin{equation}
\label{13}
\Bigg|\frac{dE_\gamma}{d\theta_\gamma{'}}\Bigg| = \gamma_A E_\gamma^0 ~
\sin \theta_\gamma{'}.
\end{equation}
The angles in the LS and the NR systems are related by 
\begin{equation}
\label{14}
\tan \theta_\gamma = \frac{1}{\gamma_A} \frac{\sin \theta_\gamma{'}}{(1+\cos
\theta_\gamma{'})}. 
\end{equation}
Then
\begin{equation}
\label{15}
\frac{d \theta_\gamma{'}}{d \theta_\gamma} = \frac{2\gamma_A}{(1+\gamma_A^2
\tan^2 \theta_\gamma)\cos^2 \theta_\gamma}. 
\end{equation}
Photon number conserving gives  
\label{16}
\begin{equation}
f(E_\gamma) d E_\gamma = f(\theta_\gamma{'}) d \theta_\gamma{'}. 
\end{equation}

The angular and energy distributions of the secondary photons  in the LS are
\begin{equation}
\label{17}
\frac{dP_{A^*}(\theta_\gamma)}{d\theta_\gamma} = f(\theta_\gamma{'})
\frac{d \theta_\gamma{'}}{d \theta_\gamma}, 
\end{equation}
 
\begin{equation}
 \label{18}
\frac{dP_{A^*}(E_\gamma)}{dE_\gamma}  = f(\theta_\gamma{'})
 \frac{d \theta_\gamma{'}}{dE_\gamma} = f(\theta_\gamma{'}) \frac{\Theta
 (2\gamma_A E_\gamma^0 - E_\gamma)}{\gamma_A E_\gamma^0 \sin
 \theta_\gamma{'}}, 
 \end{equation}
where  $\Theta(x) = 1$, if $x\leq 0$ and $\Theta(x) = 0$ if $x >0$. 
  
For an isotropic  photon distribution in the NR system 
\begin{equation}
\label{19}
f(\theta_\gamma{'}) d \theta_\gamma{'}  d \varphi_\gamma{'} =
\frac{1}{4\pi}\sin \theta_\gamma{'} d \theta_\gamma{'} d \varphi_\gamma{'}
\end{equation}
After integration over $\varphi_\gamma{'}$, we have 
\begin{eqnarray}
\label{20}
\frac{dP_{A^*}(\theta_\gamma)}{d \theta_\gamma}  & = & \frac{1}{2} \frac
{\sin \theta_{\gamma}{'} ~ d \theta_{\gamma}{'}}{d \theta_\gamma}, \\
\label{21}
\frac{dP_{A^*}(E_\gamma)}{dE_\gamma} & = & \frac{1}{2\gamma_A E_\gamma^0}
\Theta(2\gamma_A E_\gamma^0 - E_\gamma). 
\end{eqnarray}

For comparison the photon distribution of $\pi^0$ decay is 
\begin{equation}
\label{21a}
\frac{dP_{\pi^0}(E_{\gamma})}{dE_{\gamma}} =
\frac{1}{\gamma_{\pi^0}m_{\pi^0}}\Theta(\gamma_{\pi^0}m_{\pi^0} -
E_{\gamma}). 
\end{equation}

The angular photon distribution (\ref{20}) is 
\begin{equation}
\label{21b}
\frac{dP_{A^*}}{d\theta_\gamma} = \frac{2~ \gamma_A^2 ~
\sin\theta_\gamma}{(1 + \gamma_A^2 \tan^2\theta_\gamma)^2 \cos^3 \theta_\gamma}
\end{equation}
in the LS.

\subsection{Excitation of discrete nuclear level in quasi-elastic electron
scattering}

We follow the formalism of Ref.~\cite{21}. The cross section for
nuclear excitation by electron scattering 
\begin{equation}
\label{22}
e + A \to e{'} + A^*(\lambda^P,E_\gamma^0)
\end{equation}
is 
\begin{equation}
\label{23}
\frac{d \sigma_{A^*}}{d \Omega{'}} = \sigma_M(q) F_\lambda^2 (q). 
\end{equation}
Here we neglect the nuclear recoil effect. The equation is valid in 
the NR system 
where
the electron has an energy $\varepsilon_0$ and the  momentum transfer is 
\begin{equation}
\label{24}
q = 2 \varepsilon_0 \sin\frac{\Theta_{e{'}}}{2} 
\end{equation}
where $\Theta_{e'}$ is a scattering angle of the electron. 
The Mott cross section of the electron scattering off charge $Z_A$ is 
\begin{equation}
\label{27}
\sigma_M(q) = \left( \frac{2~ Z_A e^2}{q^2}~ \varepsilon_0 \cos(\theta_{e'}/2)\right)^2 =
\frac{\alpha^2 Z_A^2~cos^2(\theta_{e'}/2)}{(2 \varepsilon_0 \sin^2 (\theta_{e'}/2))^2}.
\end{equation}
Here  $\hbar = c =1$. 

A good approximation of the measured inelastic form factor $F_\lambda(q)$ ~\cite{21} is 

\begin{eqnarray} 
\label{29}
F_\lambda(q) & = & \frac{c_\lambda q^\lambda }{Z_A} \Bigg[ 1-\beta \Bigg(\lambda +
\frac{3}{2}\Bigg) + \frac{\beta}{4}(dq)^2\Bigg]e^{-(dq)^2/4} \\ 
\label{29a}
c_\lambda & = & \frac{N_\lambda \pi \sqrt{2\lambda + 1} d^{\lambda + 3}}{2^{\lambda
+ 1}}. 
\end{eqnarray}
The parameters for  $^{40}$Ca$^*(3^-$, 3.74 MeV) are $\lambda = 3$, $N_3 =
0.0496$~fm$^{-3}$, $d = 2.015$ fm and $\beta = 0.8$ .

The four-momenta of the
electron and nucleus are 
\begin{eqnarray}
k_e & = & (\varepsilon, p \sin\theta ~ \cos \varphi, p\sin\theta ~ \sin
\varphi, p \cos \theta),  \nonumber \\
\label{30}
k_A & = & (E_A, 0, 0, -p_A) 
\end{eqnarray}
in the laboratory system  and 
\begin{equation}
k_e{'}  =  (\varepsilon_0, \vec {p}_e{'}),  
%\label{31}
k_A{'}  =  (m_A, 0) 
\end{equation} 
in the nuclear rest system. 

Then 
\begin{equation}
\label{32} 
\varepsilon_0 = \frac{1}{m_A} (\varepsilon E_A + p~ p_A \cos \theta),
\end{equation}
\begin{equation}
\label{33}
q_{\textrm{max}}(\varepsilon_0) = 2~ \varepsilon_0, 
\end{equation}
where  $\theta$ is the polar angle of the electron in the LS with the $z~$- axis
parallel to $~\vec p_A$, the nuclear momentum. 

Equations (\ref{32}), (\ref{33}) allows us  to calculate the
cross section of quasi-elastic excitation electron scattering in the
system of the colliding nuclei.

\subsection{Nuclear excitation by electron or positron in peripheral heavy ion collisions}

In Fig.\ 1 we show  two stage process  
\begin{equation}
\label{34}
A_Z + A_Z \to A_Z^* + A_Z + e^+e^- ,
\end{equation}
where the $e^+e^-$ pair is produced by the virtual 
photons in the
nuclear electromagnetic field followed by nuclear excitation by the electron or positron.

%\bigskip

\begin{figure}[hbtp]
  \begin{center}
%    \resizebox{12cm}{7cm}{\includegraphics{fig1_gam.eps}}
    \resizebox{9cm}{6cm}{\includegraphics{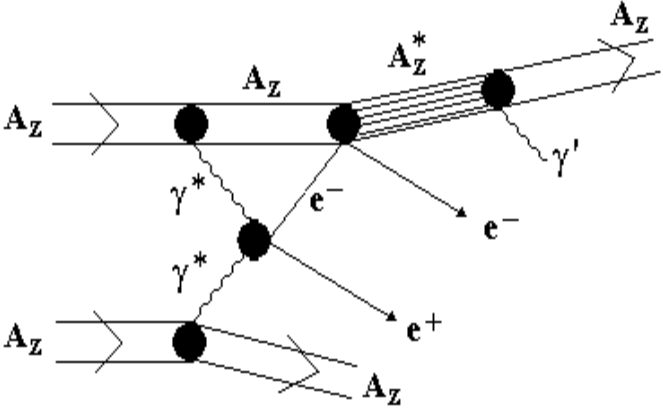}}
    \caption{Diagram of two stage process of nuclear excitation by an electron (positron) produced by the virtual photon interaction in an AA collision at the LHC.}
    \label{fig:ex1}
  \end{center}
\end{figure}

The authors of Ref.~\cite{8} showed 
that the energy and angular electron (positron) distribution 
in the double equivalent photon approximation is 
\begin{equation}
\label{35} 
\frac{d \sigma_{AA}}{d \varepsilon d \Omega} = \frac{\alpha^2}{2} p
\int\limits_{\hat E/2}^\infty \frac{d \omega}{\omega^4} \frac{2 \omega - \hat
E}{\tilde E^2} n(\omega) n \Bigg(\frac{\omega \hat E}{2\omega - \tilde E}\Bigg)
 ~ g(\omega, \theta) 
\end{equation}

where $\omega$ is the energy of the virtual photon. We have 
\begin{equation}
\label{36} 
\begin{array}{rcl}
\hat E & = & \varepsilon + p \cos \theta,  \\
\tilde E & = & \varepsilon - p \cos \theta,  \\
\varepsilon^2 & = & p^2 + m_e^2,  
\end{array}
\end{equation}

$$
g(\omega, \theta)  =  \Bigg\{ \frac{2 \omega \omega{'} - m_e^2 + (\omega
\omega{'} - m_e^2) \sin^2 \theta}{m_e^2 - (\omega \omega{'} - m_e^2) \sin^2 
\theta}    
$$

\begin{equation}
\label{36b1} 
-~~\frac{2(\omega \omega{'} - m_e^2)^2 \sin^4 \theta}{[m_e^2 
+ (\omega \omega{'} - 
m_e^2)
\sin^2 \theta]^2} \Bigg\},
\end{equation}

\begin{equation}
\label{36b2} 
\omega{'}  =  \frac{\omega \tilde E}{2 \omega - \hat E},
\end{equation}

\begin{equation}
\label{36a1}
n(\omega)  =  \frac{2}{\pi}Z_A^2 \alpha \Bigg(\frac{c}{v}\Bigg)^2 \Bigg[\xi
K_0(\xi)K_1(\xi) - \frac{v^2 \xi^2}{2c^2}(K_1^2(\xi) - K_0^2(\xi))\Bigg],   
\end{equation}

\begin{equation}
\label{36a2}
\xi =  \frac{\omega R_{\textrm{min}}}{\gamma v}.
\end{equation}

The equivalent electron approximation ~\cite{3} is more exact,
 but the results of that approach are not numerically differ from ~\cite{8}.

We consider a model of the two independent simultaneous processes in Fig.\ 1.
The cross section of the two-stage process (\ref{34}) is written as a
convolution of Eq. (\ref{35}) with the probability of nuclear
excitation,
\begin{equation}
\label{37}
\sigma_{AA^*} = \int d \varepsilon d \Omega d \Omega'
\frac{d\sigma_{AA}}{d\varepsilon d\Omega}(\varepsilon,\theta)~
P_{A^*}(q(\theta';\varepsilon,\theta)). 
\end{equation} 
We define the probability of nuclear excitation $P_{A^*} (q)$ as a ratio of
the cross section (\ref{23}) to the  sum  over all nuclear final
states
\begin{equation}
\label{38}
P_{A^*}(q) = \frac{d\sigma_{A^*}/d\Omega'}{d\sigma_{\textrm{sum}}/d\Omega'}. 
\end{equation}

We follow the Glauber model \cite{Glauber} to calculate this sum using the 
property of  closure. For electron scattering, we find  
\begin{equation}
\label{39}
\frac{d\sigma_{\textrm{sum}}}{d\Omega'} = [Z_A (Z_A-1)~ S_A^2 (q) + Z_A]~
|f_{ep}(q)|^2 
\end{equation}
where $f_{ep}(q)$ is the elastic electron-proton scattering amplitude and $S_A(q)$ is the elastic nuclear form-factor. 
For our two-stage peripheral process where 
 $q \ll 1/R_A$, ~ \\ $S_A(q) \approx 1$ so that 
\begin{equation}
\label{39a}
\frac{d\sigma_{\textrm{sum}}}{d\Omega'}(q) \approx Z_A^2 ~ |f_{ep}(q)|^2 =
\sigma_M(q).
\end{equation} 

Thus the integrated cross section, Eq. (\ref{37}), is  
\begin{eqnarray}
\label{39b}
\sigma_{AA^*} & = & \int d\varepsilon d\Omega d\Omega'
\frac{d\sigma_{AA}}{d\varepsilon d\Omega} (\varepsilon, \theta) F_\lambda^2
(q(\theta';\varepsilon,\theta))  \nonumber \\
& = & 2\pi^2 ~ \alpha^2 \int\limits_{\hat E/2}^\infty d\omega
\int\limits_{m_e}^\infty d \varepsilon \int d \cos\theta \int d \cos \theta'
 \nonumber \\
&  \times & p(\varepsilon) \frac{2\omega -\hat E}{\omega^4\tilde E^2} n(\omega) n(\omega ')
g(\omega, \theta) F_\lambda^2(q(\theta', \varepsilon, \theta))
\end{eqnarray}
where $q$ is defined  in Eq. (\ref{24}).

After  electro-excitation 
at high energies, the nucleus $A_Z^*$ essentially retains its direction.
Then the angular and energy distributions of the secondary photon are 
\begin{eqnarray}
 \label{40}
\frac{d\sigma_{AA^*}(\theta_{\gamma{'}})}{d\theta_{\gamma{'}}}
 = \sigma_{AA} ~ \frac{dP_{A^*}(\theta_{\gamma{'}})}{d\theta_{\gamma{'}}}, \\
 \label{40a}
\frac{d\sigma_{AA^*}(E_{\gamma{'}})}{dE_{\gamma{'}}}
 = \sigma_{AA} ~ \frac{dP_{A^*}(E_{\gamma{'}})}{dE_{\gamma{'}}}, 
\end{eqnarray}
where $\sigma_{AA}$ is given in Eq. (\ref{39b}). The functions
$dP_{A^*}/d\theta_\gamma$ and 
$dP_{A^*}/dE_\gamma$ depend on the angular distribution of the
  $\gamma{'}$ in the
NR system. For an isotropic $\gamma'$ distribution they are given by Eqs. (\ref{20}) 
and (\ref{21}).  

\subsection{Nuclear excitation by the peripheral strong interaction} 

One possible background for process (\ref{4})  is a  strong interaction in which one of nucleus
is excited to a discrete level
\begin{equation}
\label{41}
A_1 + A_2 \to A_1^* + A_2. 
\end{equation}
We make an rough estimate of the cross section in  peripheral
collisions with $b > (R_1 + R_2)$. 
We write the differential cross section for  elastic scattering, $A_1 +
A_2 \to A_1 + A_2$~\cite{22}, neglecting  distortion effects,  
\begin{equation}
\label{42}
\frac{d\sigma_{A_1A_2}}{dq^2} =
\frac{d\sigma_{NN}}{dq^2}(q^2)|A_1A_2S_{A_1}(\vec q) S_{A_2}(- \vec q)|^2. 
\end{equation}
Here ~$S_A(\vec q)$ ~is the nuclear ~form-factor, equal ~to ~unity 
~at ~$q=0$,
$S_A(0) = 1$. We can approximate it  simply by a Gaussian nuclear density, 
\begin{equation}
\label{43}
S_{A_i}(q) = e^{-B_{A_i} q^2/2}. 
\end{equation}
We use the standard form (see ~\cite{2}) for 
nucleon nucleon scattering cross section 
\begin{equation}
\label{44}
\frac{d \sigma_{NN}}{dq^2} (q^2) = \Bigg|
\frac{\sigma_{\textrm{tot}}(NN)}{4\sqrt{\pi}}e^{-B_Nq^2/2}\Bigg|^2. 
\end{equation}

We calculate the energy of an $AA$ collision at the LHC with $\gamma = 3500$ in
the nuclear rest system, where $\gamma_A = 2 \gamma^2 - 1$. The energy
of an $NN$-collision in the rest system is  $E_p = 2.5 \times 10^7$ GeV. We
find  \\ $\sigma_{\textrm{tot}}(NN) \approx 100$ mb and $B_{NN} \simeq 20$ GeV$^{-2}$
 at LHC energies  from a good approximation of 
 $\bar p p$ data at $\sqrt{s_{\textrm{pp}}} < 540$ GeV in Ref.~\cite{23}.

For the elastic scattering we can write the form-factor $S_A(\vec q)$ as 
\begin{equation}
\label{45}
S_A(q)  =  2\pi \int\limits_0^\infty bdbJ_0(qb)T_A(b),~~~~
T_A(b)  =   \int\limits_{-\infty}^\infty dz \rho_A(\vec b, z), \nonumber
\end{equation}
where $J_0(qb)$ is a Bessel function and $\rho_A(\vec b, z)$ is the nuclear
density.
Using Eq.~(\ref{43}), $T_A(b)$ is
\begin{equation}
\label{45a}
T_A(b) = \frac{1}{2\pi B_A} e^{-b^2/(2B_A)}.
\end{equation}
Peripheral collisions will correspond to the integral 
(\ref{45}) 
in the region $b > R$ 
\begin{equation}
\label{46}
S_A(q, R) = 2 \pi \int\limits_R^\infty bdbJ_0(qb)T_A(b). 
\end{equation}

The amplitude of  proton inelastic scattering with  nuclear excitation to
a state with spin $\lambda$ and projection $\mu$ in the same approximation
is 
\begin{equation}
\label{47}
{\cal F}_{pA^*} (\vec q) = A ~ f_{NN}(\vec q) G_{\lambda \mu}(\vec q), 
\end{equation}
where 
\begin{equation}
\label{48}
G_{\lambda \mu}(\vec q) = \int d^3 \vec r e^{i\vec q\cdot \vec r} \rho_{\lambda \mu}
(\vec r). 
\end{equation}

The inelastic  nuclear transition density is 
\begin{equation}
\label{49}
\rho_{\lambda \mu} (\vec r) = \frac{1}{A} \langle \lambda \mu
|\sum\limits_{j=1}^A \delta (\vec r - \vec r_j)|00 \rangle. 
\end{equation}
It is simple to show that $F_{\lambda \mu} (q)  |_{q \to 0} \sim q^\mu$. We define the inelastic nuclear form-factor $S_{A^*}^{(\lambda)} (q)$ as
\begin{equation}
\label{49a}
(S_{A^*}^{(\lambda)} (q))^2 = \sum\limits_\mu |G_{\lambda \mu}(\vec q) |^2. 
\end{equation}

If we assume that the state with $\mu = \lambda$ is dominant, we can 
approximate the inelastic form-factor $S_{A^*}^{(\lambda)} (q)$ by Eq. 
(\ref{29}). This form agrees
 with form-factor from inelastic electron scattering~\cite{21}. We assume
 that the charge and nucleon transition densities are the
same  for a rough estimate. We take the parameters of the 
inelastic form-factor 
for $^{40}\mbox{Ca}^*(\lambda\mu)$ from Ref.~\cite{21} and 
renormalize the form-factor 
to $Z_A/A$ so that  
\begin{equation}
\label{49d'} 
S_{A^*}^{(\lambda)}(q) = \frac{Z_A}{A}F_\lambda(q). 
\end{equation}

The Fourier transformation analogous to Eq.~(\ref{46}) is
\begin{equation}
\label{49c}
S_{A^*}^{(\lambda)}(q, R) = 2\pi \int\limits_R^\infty bdb J_\lambda (qb)
T_{A^*}^{(\lambda)}(b), 
\end{equation}
 in peripheral collisions with  $b>R$ 
where the function $T_{A^*}^{(\lambda)}(b)$ can be calculated by the inverse
Fourier transform from $S_{A^*}^\lambda (q)$ to $T_{A^*}^\lambda(b)$ 
\begin{equation}
\label{49d}
T_{A^*}^{(\lambda)}(b) = \frac{1}{2\pi} \int\limits_0^\infty qdq J_\lambda (qb)
S_{A^*}^{(\lambda)}(q). 
\end{equation}

We can calculate (\ref{49d}) analytically with the form-factor parameterisation
in Eq.~(\ref{49d'}). We have 
$$
\label{49d''}
T_{A^*}^{(\lambda)}(b)  =  \frac{1}{A}\frac{c_\lambda}{2}\Bigg[1 - \beta
\Bigg(\lambda + \frac{3}{2}\Bigg) + \beta \Bigg(\frac{b}{d}\Bigg)^2 \Bigg]
$$
\begin{equation}
\times \Bigg(\frac{4}{d^2}\Bigg) \frac{\lambda +2}{2}
\bigg(\frac{b^2}{d^2}\Bigg)^{\lambda/2} 
e^{-b^2/d^2}. 
\end{equation}

Thus Eq.~ (\ref{42}) becomes 
\begin{equation}
\label{49e}
\frac{d\sigma_{A_1^* A_2}}{dq^2} =
\frac{d\sigma_{NN}}{dq^2}(q^2)|A_1A_2S_{A_1^*}^{(\lambda)}(\vec q, R) ~
S_{A_2}(-\vec q, R)|^2. 
\end{equation}

The energy and angular distributions of the secondary photons are given by the
convolutions: 
\begin{eqnarray}
\label{49f}
\frac {d\sigma_{A_1^*A_2}} {dE_{\gamma_1{'}}} & = & \int d^2q
\frac{d\sigma_{A_1^*A_2}(q)}{dq^2} 
\frac{dP_{A_1^*}}{dE_{\gamma_1{'}}}
(E_{\gamma_1{'}}, \theta_{A_1^*}); \\
\label{49g}
\frac {d\sigma_{A_1^*A_2}} {d\theta_{\gamma_1{'}}} & = & 
\int d^2q \frac{d\sigma_{A_1^*A_2}(q)}{dq^2} 
\frac{dP_{A_1^*}}{d\theta_{\gamma_1{'}}}
(|\theta_{\gamma_1{'}} - \theta_{A_1^*}|).
\end{eqnarray}
For symmetric collisions with $A_1 = A_2 = A$ 
and  small angles,  $\theta_{A^*} = q/p_A$.

\section{Energy and angular distributions of secondary photons}

\bigskip 

We consider the peripheral processes (\ref{1a}),(\ref{4}) and (\ref{6}), 
which give us the secondary
photons. All three processes preserve the charge $Z$ and nuclear number $A$ of
the nucleus. The nucleus is excited and then emits the photons in  processes
(\ref{4}) and (\ref{6}). The $\pi^0$-meson produced in  process (\ref{1a}) by  virtual photon
fusion decays to two photons . 

We study the kinematic dependence of  $E_\gamma$ on 
$\theta_\gamma$ in the LS  for 
$^{40}$Ca (see Fig.\ 2). The beam luminosity  is  $L =
(2-4) \times
 10^{30}$ cm$^{-2}$ s$^{-1}$ (see ~\cite{2}). The
characteristics of Ca discrete level are well known~\cite{16}. In Fig.\ 2 we
use five strongly  excited levels by electrons~\cite{21}. The level
$\lambda^P = 3_1^-$, $E_\gamma^0 = 3.74$ MeV has the largest 
excitation intensity. 

Figure \ 2 demonstrates that  $E_\gamma$ falls quickly with 
 $\theta_\gamma$ at $\gamma_A = 3500$. The energy  is a few tens of
GeV in the solid angle $\theta < \theta_{\textrm{cross}}$. The  $3^-$
excitation  gives the secondary photons with  energies $22 - 
26$ GeV in the angular range of TOTEM.

Figure \ 3 shows the cross sections, Eq.~ (\ref{40}) and (\ref{40a}), 
of the two-stage 
process (\ref{4}) of the excitation of nucleus Ca by electron (positron)
in the virtual photon interaction (section 2.3). The angular
distribution of the secondary photons in the NR system are assumed
 to be isotropic
because we average over all initial directions of electrons that excite the
nucleus 
(see Eq.~(\ref{20})). The energy distribution in Fig.\ 3a is then uniform. \\ The
level $3^-$, $E_\gamma^0 = 3.74$ MeV gives the largest contribution to the sum of the angle distributions over other levels in Fig.\ 3b. The maximum in the angular distribution 
is 
$\theta_\gamma = 150$ $\mu$rad. 

The integrated cross section of the two-stage \\ process (\ref{4}) 
is equal to $\sigma_1 = 4.8$ barn at $R_{\textrm{\textrm{min}}} = 1/m_e = 386$ fm.  
The cross section $\sigma_1$ is 
large  because the $\gamma^* \gamma^* \to e^+e^-$ cross section  is rather 
large, 1.5 Kb for CaCa collisions at LHC energies.
In the TOTEM angular region, the cross section  is
16$\%$ of $\sigma_1$ and in the region up to \\ $\theta_{\textrm{cross}} = 300$
$\mu$rad  is 56$\%$ of $\sigma_1$.

\begin{figure}
\resizebox{0.45\textwidth}{!}{%
  \includegraphics{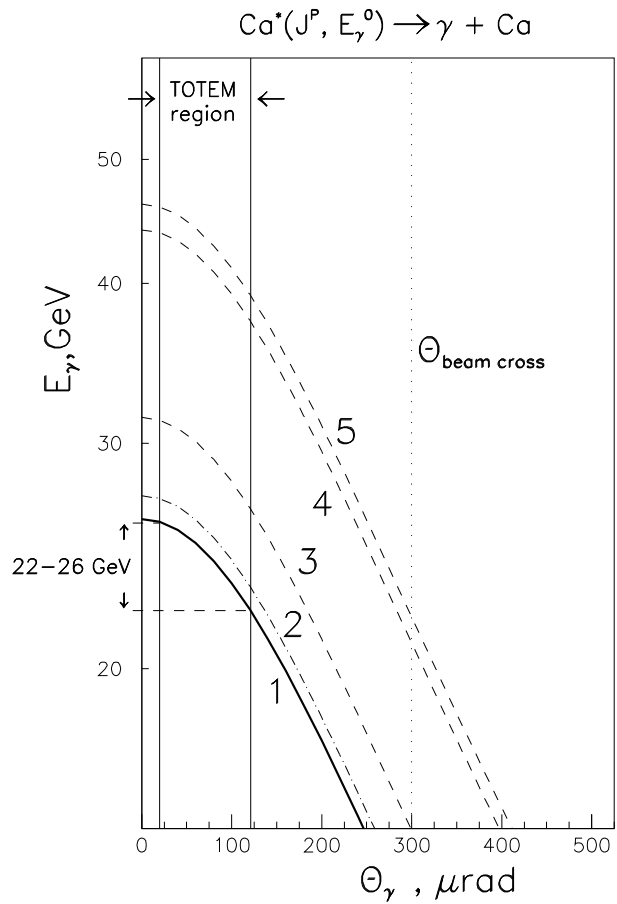}
}
\caption{ Energy  of
photons emitted by  Ca$^*(J^P,E_\gamma^0)$ as a function of angle. The lines
correspond to the discrete excited levels: (1) $3_1^-$, 3.74 MeV, (2) $2_1^+$,
3.90 MeV, (3) $5_1^-$, 4.49 MeV, (4) $3_2^-$, 6.29 MeV, (5) $3_3^-$, 6.59 MeV.
The vertical lines show the TOTEM angular region  20 $\mu$rad $\leq \theta < 120$
$\mu$rad and the  beam crossing angle.}
\label{fig2_gam}  
\end{figure}

The cross section of the process (\ref{6}), when the nucleus is 
excited by
strong peripheral nuclear interactions, is less, $\sigma_2 = 0.1$ mb,
  than the process (\ref{4}),
see Fig.\ 4. We calculate this cross section in Eq.~(\ref{49e}).
 We neglect  absorption
effects in nucleus-nucleus interactions obtaining an upper estimate. We
 also assumed that the photon angular distribution in the NR system 
is isotropic.
The calculations are for $b > R_A$ with  $R_A = 1.2A^{1/3}$ fm for
each nucleus .  

The distributions of process (\ref{1a}), photon fusion to $\pi^0$, 
 was calculated in the same way as 
our previous work~\cite{5} for PbPb collisions. The cross section of this
process is also small, $\sigma_3 = 0.19$ mb, compared with the two-stage
cross section. The energy distribution  of photons from $\pi^0$ decays 
in Fig.~4a
has a peak at $E_\gamma =0$ because we integrate over all $\pi^0$ 
direction. The angular distribution in Fig.\ 4b has a very different
dependence compared to the photon angular distribution of nuclear radiation. 

\begin{figure}
\resizebox{0.45\textwidth}{!}{%
  \includegraphics{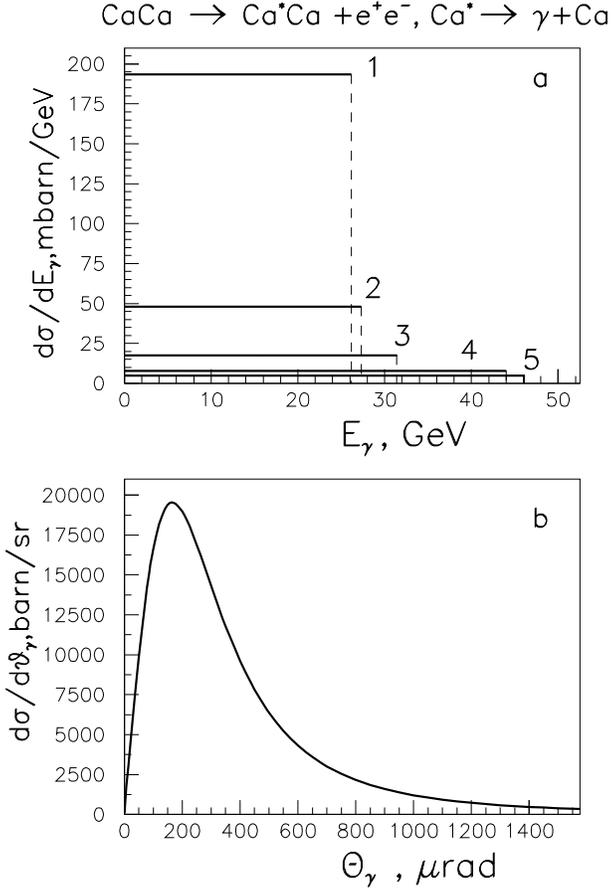}
}
   \caption{ Energy and angular distributions of secondary photons from
\break $Ca+Ca \to {\mbox{Ca}}^*(\lambda^P,E_\gamma^0)+{\mbox{Ca}} + e^+e^-$,
$Ca^* \to \gamma + {\mbox{Ca}}$. (a) Photon energy distributions, numbers 1-5, 
correspond to the five nuclear levels in Fig.\ 2. (b) Photon angular
distribution  summed over all levels.}
%label{fig3_gam}  
\end{figure}

\begin{figure}
\resizebox{0.45\textwidth}{!}{%
  \includegraphics{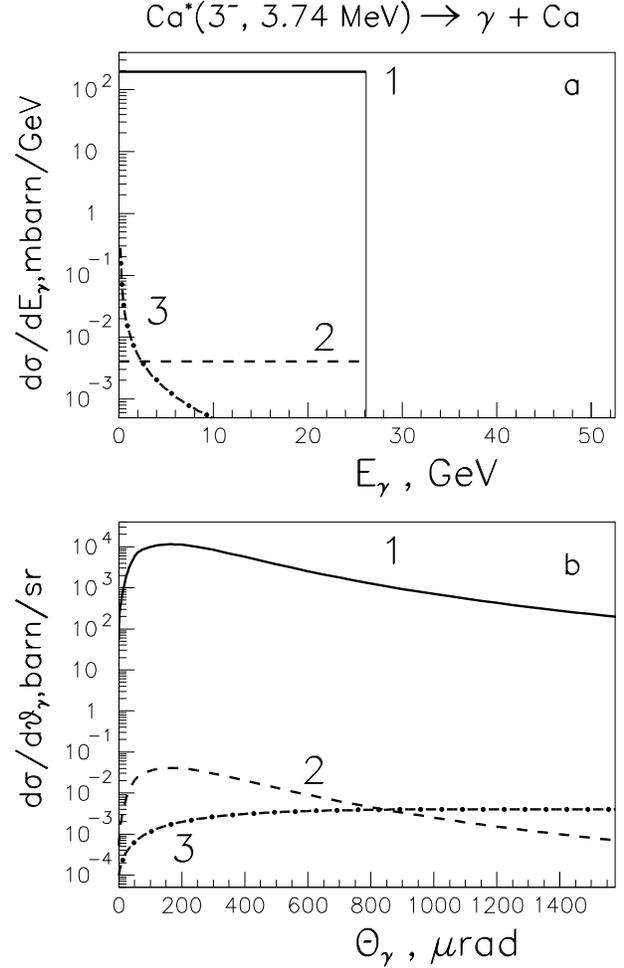}
}
  \caption{Comparison of energy (a) and angular (b) distributions of
secondary photons for three peripheral processes: (1)\ nuclear excitation by
electrons or positrons in electromagnetic interactions, (2)\  excitation
in strong interaction with $b > 2R_A$, (3)\  photons from decay $\pi^0 \to 2\gamma$ 
in process of virtual fusion in $AA$ collisions.} 
\label{fig4_gam}  
\end{figure}

\section{Conclusions}

We have presented the energy and angular distributions of  secondary photons
from nuclear radiation in  peripheral ultra-relativistic $AA$ collisions at the
LHC. The two-stage process of nuclear excitation by an electron or positron,
produced 
by  virtual photon interactions has a large cross section, $\sim 5$
b, along with a  specific angular distribution. The
secondary $\gamma$-radiation at angles $\sim 100$ $\mu$rad and at energies $\sim 20$ GeV can be measured in the region of TOTEM
Roman pots. 
High energy $\gamma$-radiation in peripheral collisions will be a
significant background in the nuclear fragmentation region with an  intensity 
of $\sim 10^6$ photons/s for CaCa collisions 
at  $L\approx 10^{30}$ cm$^{-2}$ s$^{-1}$. 

Secondary $\gamma$-radiation from nuclei can be used for to monitor the beam 
intensity,
a very difficult problem for nuclear beams. However it is
necessary to determine $R_{\textrm{\textrm{min}}}$ for nuclear
electromagnetic interaction beforehand. 

Finally secondary nuclear photons can be a good method for triggering 
processes 
(\ref{9}) or (\ref{10}) with the meson system produced in central rapidity region. The full trigger includes a signal in the central
rapidity region along with an  absence of signals from  neutrons and 
charged nuclear fragments in
nuclear fragmentation region and  high energy secondary photons  
in the Roman pots. 

We are very grateful to L. I. Sarycheva, V. A. Bodyagin, D. E. Lanskoy,
I. A. Pshenichnov and to V. V. Varlamov 
  for the useful  discussions and N. P. Karpinskaya for
 help with a manuscript. We also thank G.Baur, K. Hencken and R. Vogt for   
helpful comments on our first draft.


\end{document}